\documentclass[preprint,aps,draft]{revtex4}

\usepackage{graphicx}% Include figure files
\usepackage{dcolumn}% Align table columns on decimal point
\usepackage{bm}% bold math

\begin{document}

\setlength{\unitlength}{1mm}
\textwidth 15.0 true cm
\textheight 22.0 true cm
\headheight 0 cm
\headsep 0 cm
\topmargin 0.4 true in
\oddsidemargin 0.25 true in

\title{Elastic Inflation}

\author{Andrei Gruzinov}
 \affiliation{Center for Cosmology and Particle Physics, Department of Physics, New York University, NY 10003}

\date{April 30, 2004}

\begin{abstract}

Inflation of a universe filled by an elastic continuous medium is considered. Elastic inflation, while capable of describing current observations, is qualitatively different from standard inflationary models. The scalar and tensor modes keep evolving after crossing the horizon. Due to this superhorizon evolution, the amplitude of the tensor mode (inflationary gravitational wave) is not simply proportional to the energy density of the universe at the time of horizon crossing. The spectral index of tensor modes can be positive.

\end{abstract}


\maketitle

\section{Introduction}

The cosmological theory, and cosmological perturbation  theory in particular \cite{lif,mukh}, have been confirmed observationally only when applied to an ideal fluid. Yet inflation and inflationary perturbations are calculated assuming that the universe was dominated by a scalar field \cite{standard1,standard2,mukh}, or a higher-derivative gravity \cite{star}, a scalar field with non-minimal kinetic term \cite{k1}, a higher-derivative scalar field \cite{gho}, a self-coupled vector field \cite{v}, or by a self-coupled three form \cite{form}.

Since physics at inflationary energies is unknown, it makes sense to consider a fluid-like description of inflation, which may ultimately turn out to be closer to reality than any of the above-listed field models. An ideal fluid cannot drive inflation, but an ideal elastic medium can. Elastic medium was already considered as a model of dark energy \cite{buch}. 

We describe the relativistic elastic medium in \S2, and the elastic inflation in \S3, where we also introduce the speeds of longitudinal and transverse sound waves. We calculate the inflationary tensor spectrum in \S4, scalar spectrum in \S5, and discuss the results in \S6. 

\section{Elastic Medium}
Continuous elastic medium is a collection of time-like world lines filling the space-time (``congruence'', see \cite{buch} and references therein). The action of elastic medium is given by a time integral of potential energy along each of the world lines of the congruence. The potential is an arbitrary function of spatial distances between neighboring world lines.

It is convenient, at least for our purposes, to choose spatial coordinates coincident with the markers of the world lines, so that the world lines of the congruence become the lines of constant $x^i$, $i=1,2,3$. Then the action of the elastic medium can be written in the following form
\begin{equation}
S=-\int d^4x\sqrt{-g}V.
\end{equation}
Here $V$ is an arbitrary function of the invariants of the spatial metric 
\begin{equation}
\gamma _{ij}=-g_{ij}+{g_{0i}g_{0j}\over g_{00}}.
\end{equation}
As the three invariants, one may take 
\begin{equation}
t_1=\gamma _{ii},~~~t_2=\gamma_{ij}\gamma_{ji},~~~~~~\gamma =\det (\gamma _{ij}).
\end{equation}
We will not go beyond the first-order perturbation theory (will not calculate the non-Gaussianities for elastic inflation). Then, given that deviations from isotropy are small and $\gamma _{ij}$ is approximately diagonal, to sufficient accuracy, we may express the third invariant as a function of the other two. We choose the traces $t_1$ and $t_2$ as the variables of elastic energy, $V=V(t_1,t_2)$. All the formulas written below are valid only up to first order in metric perturbation.

The energy-momentum tensor of elastic medium is obtained by varying the action with respect to $g_{\mu \nu}$. (The non-covariant form of the following expressions is due to the special choice of coordinates. All our final results will be expressed in a coordinate-free language.)

\begin{equation}
T^{00}=Vg^{00},\label{T1}
\end{equation}
\begin{equation}
T^{0i}=Vg^{0i}+2V_1{g_{0i}\over g_{00}}-4V_2{g_{0k}g_{ki}\over g_{00}},
\end{equation}
\begin{equation}\label{T2}
T^{ik}=Vg^{ik}-2V_1\delta _{ik}+4V_2g_{ik}.
\end{equation}
Here and in the following, $V_1\equiv \partial _{t_1}V$, $V_2\equiv \partial _{t_2}V$. With $8\pi G=1$, the Einstein equations are 
\begin{equation}
G^{\mu }_{\nu }=T^{\mu }_{\nu }.
\end{equation}
When the energy-momentum tensor is known, one calculates the unperturbed inflation and inflationary perturbations along standard lines \cite{mukh}.

\section{Elastic Inflation}
The background metric is flat FRW
\begin{equation}
ds^2=a^2(\eta )\left( d\eta ^2-\delta _{ij}dx^idx^j \right),
\end{equation}
and the corresponding Friedmann equations are
\begin{equation}\label{F1}
3a^{-2}{\cal H}^2=V~~~\equiv \epsilon,
\end{equation}
\begin{equation}\label{F2}
a^{-2}(2{\cal H}'+{\cal H}^2)=V+2a^2V_1+4a^4V_2~~~\equiv -p,
\end{equation}
here and in what follows ${\cal H}\equiv a'/a$ and the prime denotes $\partial _{\eta }$. We have also defined the unperturbed energy density and pressure.

Inflationary stage requires $p\approx -\epsilon$. On the other hand, we will show that 
\begin{equation}\label{dpde}
{dp\over d\epsilon} =c_s^2-{4\over 3}c_v^2,
\end{equation}
where $c_s$ is the speed of scalar (longitudinal) sound and $c_v$ is the speed of vector (transverse) sound. Thus, inflation requires 
\begin{equation}\label{cond}
c_s^2-{4\over 3}c_v^2\approx -1, ~~~0<c_s^2<1, ~~~0<c_v^2<1.
\end{equation}
These requirements constrain inflationary potentials $V(t_1,t_2)$.

In the above equations, $c_s$ and $c_v$ are the short-wavelength limits of the propagation speeds of scalar and vector perturbations. These can be calculated neglecting gravity (assuming a pure gauge metric and neglecting the expansion of the universe). As shown in Appendix A, 
\begin{equation}\label{cs}
c_s^2=-~{3V_1+10a^2V_2+2a^2V_{11}+8a^4V_{12}+8a^6V_{22}\over V_1+2a^2V_2},
\end{equation}
\begin{equation}\label{cv}
c_v^2=-~{V_1+4a^2V_2\over V_1+2a^2V_2},
\end{equation}
where $V_{11}$,...are second derivatives. Equation (\ref{dpde}) follows.

For $c_v=0$, one can show that $V=V(\gamma )$. Then the energy depends only on the change of volume, and the elastic model reduces to an ideal fluid. We will assume that elastic inflation ends when $c_v$ vanishes. If at this moment also $c_s=0$, then one needs to add reheating. But reheating is not necessary for elastic inflation, one may assume that at the end of inflation $c_s^2=1/3$ -- inflation ends when an ultrarelativistic fluid loses elasticity.

A simple elastic inflation model, which we use for numerical examples, assumes constant sound speeds $c_{si}$, $c_{vi}$ during the inflationary epoch, satisfying $c_{si}^2-(4/3)c_{vi}^2= -1$. Inflation ends at the energy density $\epsilon _r$. After inflation, for $\epsilon < \epsilon _r$, we also assume constant sound speeds, $c_{sr}$, and $c_{vr}=0$. Equation (\ref{dpde}) gives the pressure
\begin{equation}
p=c_{sr}^2\epsilon, ~~~\epsilon < \epsilon _r,~~~~~~~~~p=(1+c_{sr}^2)\epsilon _r-\epsilon, ~~~\epsilon > \epsilon _r.
\end{equation}
Friedmann equations give energy
\begin{equation}
\epsilon = \epsilon _r+3(1+c_{sr}^2)\epsilon _rN,
\end{equation}
as a function of the number of e-foldings until the end of inflation, $N=-\ln a+{\rm const}$. We will see that this artificial ``constant speeds'' model satisfies all current observational constraints on inflation.

\section{Tensor modes}
We start with tensor modes, because the calculation of the spectrum is much simpler than for scalars. The perturbed metric is
\begin{equation}
ds^2=a^2\left( d\eta ^2-(\delta _{ij}+h_{ij})dx^idx^j\right),
\end{equation}
where $h_{ij}$ is a symmetric pure tensor, with $h_{ii}=0$ and $\partial _jh_{ij}=0$. Equations (\ref{T1}-\ref{T2}) give the energy-momentum tensor perturbation 
\begin{equation}
\delta T^i_j=2a^2(V_1+4a^2V_2)h_{ij}.
\end{equation}
The perturbation of the  Einstein tensor is 
\begin{equation}
\delta G^i_j=-{1\over 2}a^{-2}\left( h_{ij}''+2{\cal H}h_{ij}'+k^2h_{ij}\right),
\end{equation}
where $k$ is the comoving wavenumber. Dropping the indices, we get the following tensor mode equation
\begin{equation}
h''+2{\cal H}h'+k^2h+2c_v^2a^2(\epsilon +p)h=0.
\end{equation}
It is convenient to write the mode equation using cosmic time, $dt=ad\eta$. Denoting the time derivative by the dot and introducing the Hubble constant $H\equiv \dot{a}/a$, we have\begin{equation}\label{tensormode}
\ddot{h}+3H\dot{h}+{k^2\over a^2}h+2c_v^2(\epsilon +p)h=0.
\end{equation}
For zero elasticity, $c_v=0$, the above equation reduces to the standard one \cite{mukh}.

Denote by $F_T(k)$ the elastic transfer function for tensor modes defined as the large time limit of the ratio of the amplitudes of the mode described by equation (\ref{tensormode}) and the mode described by (\ref{tensormode}) with the same $a(t)$ but with $c_v=0$. The late-time ratio of amplitudes is calculated for the modes with equal early-time ($Ha\ll k$) amplitudes.

The transfer function can be calculated analytically assuming slow-roll inflation. The effect of elasticity remains small before horizon crossing and leads to a slow roll of $h$ after horizon crossing, according to
\begin{equation}
3H\dot{h}+2c_v^2(\epsilon +p)h=0.
\end{equation}
The last equation gives $h\propto F_T(k)$ with 
\begin{equation}\label{analT}
F_T(k)\approx ~\exp \left( -~{2\over 3}\int _{\epsilon _r} ^{\epsilon _k} ~{c_v^2d\epsilon \over \epsilon}~\right)
\end{equation}
Here $\epsilon _r$ is the energy density at the end of inflation, $\epsilon _k$ is the energy density at the time when the $k$ mode crosses horizon, that is when $Ha=k$. 

The slow-roll condition $c_v^2(\epsilon +p)\ll \epsilon$ will be violated near the end of inflation, leading to a multiplicative k-independent error of order few in expression (\ref{analT}). The correct expression for the transfer function is 
\begin{equation}
F_T(k)~=~F_T~\exp \left( -~{2\over 3}\int _{\epsilon _r} ^{\epsilon _k} ~{c_v^2d\epsilon \over \epsilon}~\right)
\end{equation}
where $F_T$ is a constant factor of order few.

The tensor perturbation spectrum is (here and for the scalar spectrum we use the definitions used in \cite{k2}) 
\begin{equation}\label{Ph}
P^h(k)={128\over 3}~F_T^2~{\epsilon _k\over M_P}~\exp \left( -~{4\over 3}\int _{\epsilon _r} ^{\epsilon _k} ~{c_v^2d\epsilon \over \epsilon}~\right)
\end{equation} 
The tensor spectral index (the tilt) is
\begin{equation}\label{nT}
n_T~\equiv ~{d\ln P^h(k)\over d\ln k}~=~(4c_v^2-3)\left( 1+{p\over \epsilon }\right).
\end{equation} 
For zero elasticity, $c_v=0$, this reduces to the standard expression \cite{k2}. However, elastic inflation requires $c_s^2-{4\over 3}c_v^2\approx -1$, giving 
\begin{equation}
n_T~=~3c_s^2\left( 1+{p\over \epsilon }\right).
\end{equation} 
For elastic inflation the tensor index can be positive.

For the constant speeds model of elastic inflation described at the end of \S3, we get the tensor power
\begin{equation}
P^h(k)={128\over 3}~F_T^2~{\epsilon _k\over M_P}~\left( {\epsilon _k\over \epsilon _r}\right)^{-{4\over 3}c_{vi}^2}
\end{equation} 
and the tilt
\begin{equation}
n_T~=~{c_{si}^2\over N_k}
\end{equation} 

\section{Scalar Modes}
Generic scalar perturbations of the metric are, \cite{mukh}, 
\begin{equation}
ds^2=a^2\left(~ (1+2\phi )d\eta ^2~-~2\partial _iBd\eta dx^i~-~(\delta _{ij}-2\psi \delta _{ij}+2\partial _i\partial _jE)dx^idx^j~\right) .
\end{equation} 
For this metric one finds the perturbed energy-momentum tensor from (\ref{T1}-\ref{T2}) and the perturbed Einstein tensor from \cite{mukh}. We do this in Appendix B, here we just give a summary of results. Define a new variable $\zeta$ by
\begin{equation}
-\left( 1+4{c_v^2{\cal H}^2\over k^2} \right)~\zeta~=~\Psi~+~{{\cal H}\over {\cal H}^2-{\cal H}'}\left(\Psi '+(1-{4\over 3}c_v^2){\cal H}\Psi \right),
\end{equation} 
where $\Psi \equiv \psi -{\cal H}(B-E')$ is the gauge-invariant potential. After inflation, $c_v=0$, $\zeta$ becomes the standard curvature perturbation (on superhorizon scales, for $c_v=0$, $\zeta$ is the perturbation of the scale factor on uniform energy hypersurfaces). Our $\zeta$ also coincides with the usual $\zeta$ at very early time when $k\gg {\cal H}$, allowing us to use standard results when calculating quantum fluctuations.

The $\zeta$-perturbation satisfies 
\begin{equation}\label{zmode}
\ddot{\zeta } ~+~{\dot{z}\over z}\dot{\zeta }+{c_s^2k^2\over a^2}\zeta +w\zeta =0.
\end{equation} 
Here 
\begin{equation}
z~\equiv ~{a^3(\epsilon +p)\over c_s^2H^2}
\end{equation} 
coincides with the corresponding function for an ideal fluid case \cite{k2}. The other function
\begin{equation}
w~\equiv ~ 4c_v^2H\left(2{\dot{c_v} \over c_v}-2{\dot{c_s} \over c_s}-{1\over 2}{\dot{\epsilon} \over \epsilon}\right)
\end{equation} 
turns to zero in the ideal fluid limit.

The scalar perturbation spectrum is calculated from (\ref{zmode}) similarly to the tensor mode case. Using \cite{k2}, we find
\begin{equation}\label{Pz}
P^{\zeta }(k)={16\over 9}~F_S^2~{\epsilon _k\over M_P}~{1\over c_s(1+p_k/\epsilon _k)}~\exp \{ -~{4\over 3}\int _{\epsilon _r} ^{\epsilon _k} ~{c_v^2d\epsilon \over \epsilon}\left( 1+4{d\ln c_s\over d\ln \epsilon } -4{d\ln c_v\over d\ln \epsilon }\right) ~\}.
\end{equation} 
The scalar tilt is 
\begin{equation}\label{nS}
n_S=-3\left( 1+{p\over\epsilon}\right) ~\left (~2-{d\ln c_s\over d\ln \epsilon }-{d\ln (p+\epsilon )\over d\ln \epsilon }-{4\over 3}c_v^2\left( 1+4{d\ln c_s\over d\ln \epsilon } -4{d\ln c_v\over d\ln \epsilon }\right) ~\right).
\end{equation} 

For the constant speeds inflation described at the end of \S3, the scalar power is 
\begin{equation}
P^{\zeta }(k)={16\over 9}~F_S^2~{\epsilon _k\over M_P}~{3N_k\over c_{si}}~\left( {\epsilon _k\over \epsilon _r}\right)^{-{4\over 3}c_{vi}^2}
\end{equation} 
and the tilt is 
\begin{equation}
n_S=~-~{1-c_{si}^2\over N_k}.
\end{equation} 

\section{Discussion}
Elastic Inflation is capable of describing available observations. Currently, there exist three inflationary observables \cite{teg,kom}: the tensor to scalar ratio, the scalar tilt, and non-Gaussianity (all upper bounds). The scalar power, which has been measured to a great accuracy, simply gives the energy scale of inflation, and does not constrain phenomenological models like ours.

We have not calculated the non-Gaussianity, but one expects it to be either $\sim 1/N$ (\cite{mald}) or $\sim 1$ (\cite{post}), easily passing the observational constraints. The other two constraints are satisfied even by the artificial constant speeds model of elastic inflation described at the end of \S3. This model gives a negative scalar tilt $-(1-c_{si}^2)/N$, which for $N\sim 60$ is consistent with observations for any $0<c_{si}<1$. The tensor to scalar ratio $\sim 8c_{si}/N$ is also consistent with current observations for arbitrary $0<c_{si}<1$.

Summarizing: 

(i) Elastic Inflation is described by equations (\ref{F1}, \ref{F2}, \ref{dpde}, \ref{cond}). 

(ii) The tensor and scalar spectra generated by elastic inflation are given by equations (\ref{Ph}, \ref{nT}, \ref{Pz}, \ref{nS}), where the constant factors of order unity $F_{T,S}$ should be calculated by numerical integration of the mode equations (\ref{tensormode}, \ref{zmode}).

\begin{acknowledgments}
I thank Jose Blanco-Pillado, Gia Dvali, Gregory Gabadadze, and Roman Scoccimarro for useful discussions. This work was supported by the David and Lucile Packard Foundation.
\end{acknowledgments}

\begin{appendix}

\section{Sound Speeds}

We consider perturbations on Minkowski background $ds^2=dt^2-\delta _{ij}dx^idx^j$. For short-wavelength perturbations we neglect gravity, meaning that the metric perturbation is a pure gauge, $x^i\rightarrow x^i-\xi ^i$, giving the perturbed metric
\begin{equation}
g_{00}=1,~~~~~g_{0i}=-\dot{\xi }_i,~~~~~~g_{ij}=-\delta _{ij}-\partial _i\xi _j-\partial _j\xi _i.
\end{equation} 

The perturbed traces and 4-Jacobian are
\begin{equation}
t_1=3+2\partial _k\xi _k,~~~~~~t_2=3+4\partial _k\xi _k,~~~~~~\sqrt{-g}=1+\partial _k\xi _k.
\end{equation} 
The perturbed energy-momentum tensor is calculated from (\ref{T1}-\ref{T2}):
\begin{equation}
T^0_0=V+\delta V,~~~~~T^0_i=(2V_1+4V_2)\dot{\xi }_i,~~~~~T^i_0=-2(V+2V_1+4V_2)\dot{\xi }_i,~~~~~
\end{equation} 
\begin{equation}
T^i_k=(\delta V+2\delta V_1+4\delta V_2)\delta _{ik}+(2V_1+8V_2)(\partial _i\xi _k-\partial _k\xi _i),
\end{equation} 
where $V_1\equiv \partial _{t_1}V$,... and the perturbations of potential energy and its derivatives are given by 
\begin{equation}
\delta V=(2V_1+4V_2)\partial _k\xi _k,~~~\delta V_1=(2V_{11}+4V_{12})\partial _k\xi _k,~~~\delta V_2=(2V_{21}+4V_{22})\partial _k\xi _k.
\end{equation} 

The equation of motion 
\begin{equation}
0=\nabla _\mu T^\mu _\nu ={1\over \sqrt{-g}}\partial _\mu (\sqrt{-g}T^\mu _\nu )-{1\over 2}(\partial _\nu g_{\alpha \beta })T^{\alpha \beta }
\end{equation} 
gives 
\begin{equation}
(2V_1+4V_2)\ddot{\xi }_k+\partial _k(\delta V+2\delta V_1+4\delta V_2)+(2V_1+8V_2)(\partial _n^2\xi _k+\partial _k\partial _n\xi _n).
\end{equation} 
For vector modes ($\partial _n\xi _n=0$) we get the propagation speed (\ref{cv}), with a=1. For the scalar mode ($\xi _n\equiv\partial _n\chi$), we get the speed (\ref{cs}).

\section{Scalar Perturbations}
The perturbed metric is  
\begin{equation}
ds^2=a^2\left(~ (1+2\phi )d\eta ^2~-~2\partial _iBd\eta dx^i~-~(\delta _{ij}-2\psi \delta _{ij}+2\partial _i\partial _jE)dx^idx^j~\right) .
\end{equation} 
Perturbation of the energy-momentum tensor is found from (\ref{T1}-\ref{T2}) 
\begin{equation}
\delta T^0_0=\delta V,~~~~~\delta T^0_i=(2a^2V_1+4a^4V_2)\partial _iB,
\end{equation} 
\begin{equation}
\delta T^i_k=(\delta V+2a^2\delta V_1+4a^4\delta V_2)\delta_{ik}+(4a^2V_1+16a^4V_2)(\partial _i\partial _kE-\psi \delta _{ik}).
\end{equation} 
Here the perturbations of potential energy and its derivatives are 
\begin{equation}
\delta V=V_1\delta t_1+V_2\delta t_2,~~~\delta V_1=V_{11}\delta t_1+V_{12}\delta t_2,~~~\delta V_2=V_{21}\delta t_1+V_{22}\delta t_2,
\end{equation} 
and the perturbations of the traces are 
\begin{equation}
\delta t_1=-2a^2(3\psi +k^2E),~~~~~\delta t_2=-4a^4(3\psi +k^2E),
\end{equation}
where $k$ is the comoving wavenumber.

With Einstein tensor written in \cite{mukh}, the Einstein equations are 
\begin{equation}
-3{\cal H}^2\phi-3{\cal H}\psi '-k^2\psi +k^2{\cal H}(B-E')={1\over 2}a^2\delta V,
\end{equation}
\begin{equation}\label{Ein}
{\cal H}\phi+\psi '=a^4(V_1+2a^2V_2)B, 
\end{equation}
\begin{eqnarray}
(2{\cal H}'+{\cal H}^2)\phi +{\cal H}\phi '+\psi ''+2{\cal H}\psi '-{k^2\over 2}D=\nonumber\\
-{1\over 2}a^2(\delta V+2a^2\delta V_1+4a^4\delta V_2)+2a^4(V_1+4a^2V_2)\psi ,
\end{eqnarray}
\begin{equation}
D=4a^4(V_1+4a^2V_2)E,
\end{equation}
\begin{equation}
D\equiv\phi -\psi +(B-E')'+2{\cal H}(B-E').
\end{equation}

This system is redundant. Due to time reparametrization invariance one equation can be dropped. Equivalently, one gauge condition may be imposed. We set 
\begin{equation}
B=0.
\end{equation}
Then (\ref {Ein}) gives
\begin{equation}
{\cal H}\phi +\psi '=0.
\end{equation}
Introducing a gauge invariant potential (see \cite{mukh} for discussion)
\begin{equation}
\Psi =\psi +{\cal H}E',
\end{equation}
we get a system of two first-order ODEs
\begin{equation}\label{sys1}
\Psi '+{\cal H}\Psi=a^4(V_1+2a^2V_2)E'-4a^4(V_1+4a^2V_2){\cal H}E
\end{equation}
\begin{equation}\label{sys2}
k^2\Psi =a^4(V_1+2a^2V_2)(3\Psi -3{\cal H}E'+k^2E).
\end{equation}

Now we want to get an equation for the curvature perturbation $\zeta$. More precisely, we would like to define a variable which becomes curvature perturbation in the ideal fluid limit $c_v=0$ (because this is what we want to calculate), and also coincides with the standard $\zeta$ variable at early times (this will allow to use standard results to normalize the mode). This can be done in the following (rather tedious) way.

One notes that the system (\ref{sys1}, \ref{sys2}) can be written as 
\begin{equation}
\left(2a\Psi +a^3(\epsilon +p)E\right)'=-3c_s^2{\cal H}a^3(\epsilon +p)E,
\end{equation}
\begin{equation}
k^2\left( 2a\Psi +a^3(\epsilon +p)E\right)=-3a^3(\epsilon +p)(\Psi-{\cal H}E').
\end{equation}
With the help of \cite{k2}, one defines a possible variable $\zeta$:
\begin{equation}
\zeta ={k^2\over 3}(E+{2\over a^2(\epsilon +p)}\Psi ).
\end{equation}
The definitions and equations given in the main text follow.

\end{appendix}

\end{document}